\documentclass[showpacs]{revtex4}
\usepackage{amsmath}
\usepackage{amssymb}

\begin{document}
	
%\title{	Propagation of gravitational waves in different models of dark energy}
\title{	Dispersion relations for gravitational waves in different models of dark energy}

\author{
	Vladimir Dzhunushaliev,$^{1,2,3,4}$
	\footnote{v.dzhunushaliev@gmail.com}
	Vladimir Folomeev,$^{2,3,4}$
	\footnote{vfolomeev@mail.ru}
	Burkhard Kleihaus,$^{4}$
	\footnote{b.kleihaus@uni-oldenburg.de }
	Jutta Kunz$^4$
	\footnote{jutta.kunz@uni-oldenburg.de}
}
\affiliation{
	$^1$
	Department of Theoretical and Nuclear Physics,  Al-Farabi Kazakh National University, Almaty 050040, Kazakhstan \\
	$^2$ Institute of Experimental and Theoretical Physics,  Al-Farabi Kazakh National University, Almaty 050040, Kazakhstan \\
	$^3$Institute of Physicotechnical Problems and Material Science of the NAS
	of the Kyrgyz Republic, 265 a, Chui Street, Bishkek 720071,  Kyrgyz Republic \\
	$^4$Institut f\"ur Physik, Universit\"at Oldenburg, Postfach 2503
	D-26111 Oldenburg, Germany
}

\begin{abstract}
The propagation of weak gravitational waves on the background of dark energy is studied.
The consideration is carried out within the framework of an approximate approach where the cosmological scale factor is expanded as
a power series for relatively small values of the redshift corresponding to the epoch of the present accelerated expansion of the Universe.
For several different dark energy models, we obtain dispersion relations for gravitational waves which can be used to estimate the viability
of every specific model by comparing with observational data.
\end{abstract}

\maketitle	
\section{Introduction}

The discovery of the accelerated expansion of the present Universe at the end of the 1990's forced theorists to seek  mechanisms
providing such an acceleration. Among such mechanisms perhaps the most popular is a hypothesis that the Universe is filled by
a special substance, called dark energy (DE). The true nature of the latter is still unknown, but its main properties
which enable us, on the one hand, to model the acceleration and on the other -- not to contradict the observational astronomical data, -- are quite clear.
Namely, DE has to be homogeneously and isotropically distributed on large cosmological scales and has negative effective pressure $p$
whose modulus is comparable to the energy density $\rho$, i.e., $|p|\sim \rho$.

The literature in the field offers a variety of different DE models
(see, e.g., the book~\cite{AmenTsu2010}) possessing the aforementioned properties.
Independent of the fact which of them describes the observations in the most adequate way,
the very fact of the possibility for DE to be present in the Universe assumes that all processes on the cosmological scales
must take place on the background of DE. In particular, it concerns the process of propagation of gravitational waves (GWs)
which represent small-scale spacetime perturbations. Since gravity is a universal type of interaction, during the propagation of a GW
it, in general, will inevitably interact with any types of matter in the Universe, including DE.

 Studies of interaction of GWs with matter filling the Universe are performed in the literature for a long time (see, for example, the earlier paper \cite{Grishchuk:1974ny}
 where the case of pressure $p>0$ is under consideration). After the discovery of the accelerated expansion of the present Universe
 there arose an interest in studying the process of the propagation of a GW through matter with negative pressure. In the simplest case it can be the cosmological  $\Lambda$-term
 \cite{Bernabeu:2011if,Arraut:2012xr} or matter with a linear equation of state~\cite{Dunsby:1998hd,Kleidis:2005}.
 In the present paper we consider an approximate approach where DE can be modelled in arbitrary form (fluid or scalar field),
 and the consideration of the propagation of a GW is performed on relatively small cosmological scales.
 This enables us to estimate the magnitude of the dispersion of a GW,
which is interesting from the observational point of view.
 Comparing the dispersion for different types of DE, one can obtain information with respect to the viability of specific models of DE.

\section{General equations and particular solutions}

We consider the propagation of a weak GW on the background of a spacetime filled by DE. In doing so, we follow
Ref.~\cite{Kleidis:2005}. The equation for the GW is
\begin{equation}
\label{GW_eq_gen}
	h_{\mu\nu;\alpha}^{;\alpha}-2 R_{\alpha\mu\nu\beta}^{(0)} h^{\alpha\beta}=0,
\end{equation}
where  Greek indices refer to the 4-dimensional  spacetime, $ R_{\alpha\mu\nu\beta}^{(0)}$ is the background Riemann tensor.
We use the Lorentz gauge choice
\begin{equation}
\label{gauge}
	\left(
		h^{\alpha\beta}-\frac{1}{2}g^{\alpha\beta}h
	\right)_{;\beta}=0.
\end{equation}
It is assumed that the wave propagates in a spatially flat universe described by the metric
\begin{equation}
\label{metric}
	ds^2 = dt^2-a^2(t)\left(
		\delta_{ik}+h_{ik}
	\right)dx^i dx^k,
\end{equation}
where  Latin indices refer to the 3-dimensional space, $\delta_{ik}$ is the Kronecker symbol.
Hereafter we work in  units $c=\hbar=1$,
and we employ the conformal time coordinate, defined as
$d\eta=dt/a(t)$. This leads us to the following GW equation:
\begin{equation}
\label{GW_eq_Grish}
	h_{ik}^{\prime\prime}+2\frac{a^\prime}{a}h_{ik}^{\prime}+\delta^{lm} h_{ik, l m}=0,
\end{equation}
where the prime denotes differentiation with respect to $\eta$ . Let us seek a solution to Eq.~\eqref{GW_eq_Grish} in the form
\begin{equation}
\label{GW_eq_sol}
	h_{ik}(\eta, x^j) = \frac{h(\eta)}{a(\eta)}\alpha(k)\varepsilon_{ik}e^{i k_j x^j},
\end{equation}
where $k_j$ is the comoving wave vector, $\alpha$ is the dimensionless amplitude of the GW, and $\varepsilon_{ik}$ is the
corresponding polarization tensor.  Inserting \eqref{GW_eq_sol} into \eqref{GW_eq_Grish}, we have for the time-dependent function $h$
\begin{equation}
\label{GW_eq_sol_time}
	h^{\prime \prime}+\left(
		k^2 -\frac{a^{\prime\prime}}{a}
	\right)h=0.
\end{equation}

Since we consider the propagation of a GW on the background spacetime, then, to find the function $a$,
it is necessary to solve the background Einstein equations with the given matter source. To do this, let us employ the (0-0) component of the Einstein equations
\begin{equation}
\label{Ein_00}
	\left(
		\frac{a^\prime}{a^2}
	\right)^2 = \frac{8\pi G}{3}\rho(\eta).
\end{equation}
As the matter source, here we use a perfect fluid with  $T_{\mu\nu}=\text{diag}(\rho, -p, -p, -p)$. As a result, we have two equations
 \eqref{GW_eq_sol_time} and \eqref{Ein_00}, whose solution describes
 the propagation of a GW through any type of matter given by the right-hand side of Eq.~\eqref{Ein_00}.

Below we consider several examples involving different types  of DE. Our purpose will be the derivation of a dispersion relation,
which, as one can see from Eq.~\eqref{GW_eq_sol_time},  is defined in terms of the scale factor $a$. The solution for the latter will be sought
as follows. We will consider the evolution of the Universe on the relatively small time scales, say, starting from its transition to the stage of the present
accelerated expansion, i.e., from  $z\approx 0.4$. In doing so, let us expand the scale factor as a power series in $\tau$
\begin{equation}
\label{a_ser}
	a\approx a_p+a_1 \tau+\frac{1}{2}a_2 \tau^2,
\end{equation}
where $\tau=\eta-\eta_p$, and the index $p$ refers to the value of the corresponding variable somewhere in the past. A similar expansion
will be used for the density of matter $\rho(\eta)$. Substituting these expansions into Eqs.~\eqref{GW_eq_sol_time} and \eqref{Ein_00},
algebraic equations for the expansion coefficients will be obtained.

Proceeding in this way and using \eqref{a_ser}, one can, in particular, find the following expansion in
 $\tau$ for the ratio $a^{\prime\prime}/a$ from~\eqref{GW_eq_sol_time}:
\begin{equation}
\label{a_pr_pr}
	\frac{a^{\prime\prime}}{a} \approx \frac{a_2}{a_p}\left[
		1-\frac{a_1}{a_p} \tau+\frac{1}{a_p^2}\left(
			a_1^2-\frac{1}{2}a_p \,a_2
		\right)\tau^2
	\right]. 
\end{equation}
Then the dispersion relation for a given instant of time can be written in the form
\begin{equation}
\label{disp_rel_gen}
	\omega^2=k^2 -\frac{a^{\prime\prime}}{a}
	%\omega^2=k^2 c^2-\frac{a^{\prime\prime}}{a}
\end{equation}
with $a^{\prime\prime}/a$ taken from \eqref{a_pr_pr}.

Depending on a specific DE model, we will obtain different values for the expansion coefficients
$a_1, a_2$ containing parameters  of the given DE model. Correspondingly, we will obtain different forms of the dispersion relation  \eqref{disp_rel_gen}
that will enable us (at least theoretically) to distinguish different types of DE by tracking changes in the form of a GW.
%with respect to a change in the form of a GW.

\subsection{Chaplygin gas}

As a first example of a fluid filling the Universe,
let us choose the generalized Chaplygin gas model, for which the equation  of state can be taken in the form \cite{Bento:2003we}
\begin{equation}
\label{eos_chap}
	p=-\rho_0\frac{A_s}{(\rho/\rho_0)^\alpha}, \quad \rho=\rho_0\left[A_s +
	\frac{1-A_s}{a^{3(1+\alpha)}}\right]^{\frac{1}{1+\alpha}}.
\end{equation}
Here $\rho_0$ is some ``characteristic'' density of the Universe, $\alpha$ and $A_s$ are positive constants.
The limiting case $A_s=1$ corresponds to the de Sitter universe, and the case $A_s=0$~ -- to the universe filled by dust.

Using the ansatz \eqref{eos_chap} in \eqref{Ein_00} and applying the expansion
\eqref{a_ser}, one can find the following expressions for the expansion coefficients:
\begin{equation}
\label{a_i_chap}
	a_1^2 = \frac{8\pi G}{3}a_p^4 \,\rho_0 A^{\frac{1}{1+\alpha}}, \quad
	a_2 = 4\pi G \rho_0 A^{\frac{1}{1+\alpha}}
	\left(
		\frac{4}{3}a_p^3-\frac{1-A_s}{A a_p^{4+3\alpha}}
	\right),
\end{equation}
where $A=A_s+(1-A_s)/a_p^{3(1+\alpha)}$. Note here that by equating the obtained coefficient
 $a_2$ to zero [i.e., equating the parentheses in \eqref{a_i_chap} to zero], it is possible to find a value
 $a_p^{\text{change}}$ at which the decelerated expansion is replaced by acceleration.
 In particular, using  the observational constraints on $\alpha$ and $A_s$
($0.2\lesssim \alpha \lesssim 0.6, \, 0.76\lesssim A_s \lesssim 0.88$, see Ref.~\cite{Bento:2003we}),
one can find that $a_p^{\text{change}}$ lies in the range $\sim 0.7-0.78$.
These values can be used as a lower limit on $a_p$ which approximately defines the beginning of the stage of the DE domination.
In turn, as an upper limit for
$a_p$ one can choose, for example, the current value of the scale factor $a_p^{\text{current}}=1$.
Using these values of $a_p$ and the aforementioned values of the parameters $A_s$ and $\alpha$, one can estimate the influence that
DE in the form of the Chaplygin gas has on the propagation of a GW.
%(as compared, e.g., to Minkowski spacetime for which  $a^{\prime \prime}$ in Eq.~\eqref{disp_rel_gen}
%is equal to zero).

As an example, let us write down a concrete form of the expression \eqref{a_pr_pr} [and correspondingly the dispersion relation~\eqref{disp_rel_gen}],
when \eqref{a_i_chap} is inserted. For simplicity, let us choose  $a_p=1$.
Then we have
$$
\frac{a^{\prime\prime}}{a} \approx \frac{4\pi G}{3}\rho_0\left[4-3(1-A_s)\right]\left[
1-\frac{2}{3}\sqrt{6\pi G \rho_0}\tau+2\pi G \rho_0 (1-A_s)\tau^2
\right].
$$
This expression describes the deviation, 
which appears in the dispersion relation 
due to the presence of DE in the Universe, 
in the form of the Chaplygin gas, as compared,
for example, with empty Minkowski spacetime, for which  $a^{\prime \prime}$ is equal to zero.

\subsection{Quintessence}

As a second example, consider the well-known quintessence DE model described by a canonical scalar field
$\phi$ with the potential $V(\phi)$ whose action is
%that interacts with all the other components only through
%standard gravity. Such a quintessence model
\begin{equation}
\label{quint_action}
	S=\int d x^4\sqrt{-g}\left[
		-\frac{R}{16\pi G}+\frac{1}{2}\partial_\mu  \phi \partial^\mu  \phi-V(\phi)
	\right].
\end{equation}
By varying this action with respect to the metric and the scalar field, one can obtain the corresponding Einstein and scalar-field equations,
which for the Friedmann-Robertson-Walker Universe can be written in the following form:
\begin{eqnarray}
\label{sf_eq_quint}
	\frac{1}{a^2}\left(
		\phi^{\prime\prime}+2\frac{a^\prime}{a}\phi^\prime
	\right) &=& -\frac{d V}{d\phi},
\\
\label{Einstein-00_quint}
	\left(\frac{a^\prime}{a^2}\right)^2 &=& \frac{8\pi G}{3}\left[
		\frac{1}{2 a^2}\phi^{\prime 2}+V(\phi)
	\right].
\end{eqnarray}
A description of DE within the framework of such a model is carried out for a certain choice of the quintessence potential $V(\phi)$.
For example, consider a power potential used within the framework of the so-called  ``freezing models'' \cite{Zlatev:1998tr}:
\begin{equation}
\label{poten_quint}
	V(\phi)=M^{4+n}\phi^{-n} \quad (n>0),
\end{equation}
where the free parameter $M$ is determined from the observational constraints,
with
$M\approx \left(\rho_m M_p^n\right)^{1/(n+4)}$.
Here $\rho_m\approx 10^{-47} \text{GeV}^4$
is the current matter density, $M_p$ is the Planck mass. The current value of the field $\phi$ is assumed to be of the order of $M_p$.

The solution of the system \eqref{sf_eq_quint} and \eqref{Einstein-00_quint} is sought in the form of the expansions
\begin{equation}
\label{phi_ser}
	\phi \approx \phi _p+\phi _1 \tau+\frac{1}{2}\phi _2 \tau^2
\end{equation}
and \eqref{a_ser} for the scale factor. As a result, the following expressions for the expansion coefficients can be found:
\begin{eqnarray}
\label{a_i_quint}
	a_1 &=& \pm \sqrt{\frac{4\pi G}{3}}a_p\sqrt{\phi_1^2+2 a_p^2 V(\phi_p)}, \quad
	a_2=-\frac{4\pi G}{3}a_p\left[\phi_1^2-4 a_p^2 V(\phi_p)\right],
\nonumber\\
	\phi_2 &=& n\frac{ a_p^2}{\phi_p}V(\phi_p)\mp\sqrt{\frac{16\pi G}{3}}\phi_1\sqrt{\phi_1^2+2 a_p^2 V(\phi_p)}.
\end{eqnarray}
Two other coefficients $\phi _p$ and $\phi _1$ are arbitrary. Since in the quintessence model the scalar field evolves in time,
it is obvious that at various instants   $\phi _p$ and $\phi _1$ will have different values. In particular,
when the Universe transits to the stage of the present accelerated expansion the relation between
 $\phi _p$ and $\phi _1$ can be found from the condition
  $a_2=0$ and $a_p\approx 0.7$ (the latter value follows from observations).

For the current instant of time, when  $a_p^{\text{current}}=1$,
one can get the following estimates.
%Proceeding from the condition of (necessity) obtaining
In order to obtain the late-time cosmic acceleration in the model with the potential
\eqref{poten_quint}, it is necessary to provide the following current value of the scalar field~\cite{AmenTsu2010}:
\begin{equation}
\label{phi_p_cond_quint}
	\phi _p>\frac{n}{4\sqrt{\pi}}M_p.
\end{equation}
That is,  for $n=\mathcal{O}(1)$ the field value is of the order of the Planck mass.
In addition, as in the case of the inflation in the early Universe, to obtain the accelerated expansion in the present Universe,
the field energy must be concentrated in the potential part,
i.e. we must have $\phi_1^2 \ll V(\phi_p)$. Taking all this into account,  Eq.~\eqref{a_i_quint} gives:
\begin{equation}
\label{a_1_a_2_approx_quint}
a_1\approx \pm \left(\frac{4\sqrt{\pi}}{n}\right)^{n/2}\sqrt{\frac{8\pi G}{3}\rho_m}, \quad
a_2\approx \left(\frac{4\sqrt{\pi}}{n}\right)^{n}\frac{16 \pi G}{3}  \rho_m.
\end{equation}
Using these values in Eqs.~\eqref{a_pr_pr} and \eqref{disp_rel_gen}, one can obtain a dispersion relation for the quintessence model under consideration
whose form eventually will be determined  by the current matter density $\rho_m$ and the value of the free parameter $n$.
Namely, the resulting expression can be represented in the form
$$
\frac{a^{\prime\prime}}{a} \approx \frac{16\pi G}{3}\left(\frac{4\sqrt{\pi}}{n}\right)^n \rho_m
\left[
1 \mp \sqrt{\frac{8\pi G}{3}\rho_m}\left(\frac{4\sqrt{\pi}}{n}\right)^{n/2}\tau.
\right]
$$

\subsection{$k$-essence}

Consider the case of a scalar field with a non-canonical kinetic term which is employed in
 $k$-essence models~\cite{AmenTsu2010}. In this case
the cosmic acceleration can be realized by the kinetic energy of the scalar field. In such models the scalar-field Lagrangian $L_\phi=P(\phi, X)$
is in general some function  of the scalar field $\phi$
and its kinetic energy $X=1/2 \partial_\mu  \phi \partial^\mu  \phi$.

In the literature, various forms of
$L_\phi$ are under consideration.
For our purposes, we employ one of the simplest variants where
\begin{equation}
\label{lagr_phi_k_essence}
L_\phi=P=-X+\frac{1}{M^4}X^2.
\end{equation}
This is the so-called ghost condensate model~\cite{ArkaniHamed:2003uy},
in which the Lagrangian depends only on kinetic terms.
Here $M$ is a constant with the dimension of mass.
As pointed out in~\cite{ArkaniHamed:2003uy}, by choosing  $M\sim 10^{-3}\text{eV}$,
it is possible to describe the current accelerated expansion of the Universe.

In the language of hydrodynamics, the function $P$ plays the role of pressure. In turn, the energy density
$\rho=2 X P_X-P$, where the index $X$ denotes differentiation with respect to $X$.
Using the Lagrangian \eqref{lagr_phi_k_essence}, one can obtain the following equation for the scalar field:
\begin{equation}
\label{sf_eq_k_essence}
	\left(P_X+2 X P_{X X}\right)\ddot\phi + 3\frac{\dot{a}}{a}P_X \dot \phi=0,
\end{equation}
where the dot denotes differentiation with respect to the cosmic time $t$.

Changing to the conformal time $\eta$, Eq.~\eqref{sf_eq_k_essence}
with $P$ from \eqref{lagr_phi_k_essence} and
the (0-0) component of the Einstein equations take the form:
\begin{eqnarray}
\label{sf_eq_k_essence_conf}
	\left(
		\frac{6}{M^4}X-1
	\right)\phi^{\prime\prime}-
	2\frac{a^\prime}{a}\phi^\prime &=& 0,
\\
	\left(\frac{a^\prime}{a^2}\right)^2 &=& \frac{8\pi G}{3}X\left[
		\frac{3}{M^4}X-1
	\right].
\label{Ein_00_k_essence_conf}
\end{eqnarray}
Here $X=\phi^{\prime 2}/(2 a^2)$, and the prime again denotes differentiation with respect to $\eta$ .
Then, expanding the scalar field according to \eqref{phi_ser}
and the scale factor as \eqref{a_ser}, we find the following expressions for the expansion coefficients:
\begin{equation}
\label{a_i_k_ess}
	a_1=\pm\frac{\sqrt{2\pi G}}{M^2}\phi_1\sqrt{\phi_1^2-\frac{2}{3}a_p^2 M^4}, \quad
	a_2=\frac{4\pi G}{3}a_p \phi_1^2, \quad
\phi_2=\frac{\sqrt{8\pi G}}{3}\frac{a_p M^2 \phi_1^2}{\phi_1^2-a_p^2 M^4/3}\sqrt{\phi_1^2-\frac{2}{3}a_p^2 M^4}.
	\end{equation}
As in the case of the quintessence from the previous section, the coefficient  $\phi _1$ is arbitrary.
Note that $\phi _p$ is not present in the above expressions, since we consider the case of pure kinetic $k$-essence.

Using the expansion coefficients \eqref{a_i_k_ess} in \eqref{a_pr_pr}, one can obtain the dispersion relation  \eqref{disp_rel_gen} with
$$
\frac{a^{\prime\prime}}{a} \approx \frac{4\pi G}{3} \phi_1^2\left[
1 \mp \frac{\sqrt{2\pi G}}{M^2}\phi_1 \sqrt{\phi_1^2-\frac{2}{3}M^4}\,\tau+2\pi G \phi_1^2\left(\frac{\phi_1^2}{M^4}-1\right)\tau^2
\right].
$$

\subsection{Another type of $k$-essence}

One more form of $k$-essence is the so-called dilatonic ghost condensate~\cite{AmenTsu2010} whose Lagrangian is
\begin{equation}
\label{lagr_phi_k_essence_another}
L_\phi=P=-X+\frac{e^{\kappa \lambda \phi}}{M^4}X^2,
\end{equation}
where $\kappa=\sqrt{8\pi G}$, $\lambda$ is a dimensionless constant. When $\lambda=0$,
we return to the model \eqref{lagr_phi_k_essence}. In the model
\eqref{lagr_phi_k_essence_another} the scalar-field equation and the (0-0) component of the Einstein equations are (expressed already in terms of the conformal time):
\begin{eqnarray}
\label{sf_eq_k_essence_another_conf}
	\left(
		\frac{6}{M^4}e^{\kappa \lambda \phi}X-1
	\right)\phi^{\prime\prime}-
	2\frac{a^\prime}{a}\phi^\prime +\frac{3 \kappa }{M^4} \lambda e^{\kappa \lambda \phi} X^2&=& 0,
\\
	\left(\frac{a^\prime}{a^2}\right)^2 &=& \frac{\kappa^2}{3}X\left[
		\frac{3}{M^4}e^{\kappa \lambda \phi}X-1
	\right],
\label{Ein_00_k_essence_another_conf}
\end{eqnarray}
where again $X=\phi^{\prime 2}/(2 a^2)$. Using the expansions
\eqref{phi_ser} and \eqref{a_ser}, we have the following expansion coefficients for such a model:
\begin{eqnarray}
\label{expan_coeff_k_essence_another}
	a_1 &=& \pm \frac{\kappa}{2 M^2}\phi_1\sqrt{e^{\kappa \lambda \phi_p}\phi_1^2-\frac{2}{3}a_p^2 M^4}, \quad
	a_2=\frac{\kappa^2}{6}a_p\, \phi_1^2\left[
1\mp \frac{3\left(1-a_p^2\right)}{2 a_p^3 M^2}
\lambda e^{\kappa \lambda \phi_p}  \frac{\phi_1^2}{\sqrt{e^{\kappa \lambda \phi_p}\phi_1^2-\frac{2}{3}a_p^2 M^4}}
\right],
\nonumber\\
	\phi_2 &=& \pm \frac{\kappa \phi_1^2}{3 a_p^2\left(e^{\kappa \lambda \phi_p}\phi_1^2-a_p^2 M^4/3\right)}
\left(a_p^3 M^2\sqrt{e^{\kappa \lambda \phi_p}\phi_1^2-\frac{2}{3}a_p^2 M^4}\mp \frac{3}{4}e^{\kappa \lambda \phi_p} \lambda \phi_1^2
\right).
\end{eqnarray}
When $\lambda=0$, we return to the model \eqref{lagr_phi_k_essence} with the coefficients \eqref{a_i_k_ess}.
Taking into account that when one chooses $\lambda > 0$ (as is done in the literature)
for some instant of time in the past $a_p<1$, the coefficient 
  $a_2$ can already 
%  (?in principle?) 
  become negative by choosing the upper sign in the expression for $a_2$, i.e., the minus sign.
  This assumes that  $\phi_1$ should be positive to provide  $a_1>0$
  that corresponds to the expansion of the Universe, and not to the contraction. Then, 
  equating to zero the square brackets in the expression for
$a_2$, one can find a relation between $\phi_1$ and $\phi_p$
by virtue of  $a_p\approx 0.7$ and given values of $\lambda$ (for example, $\lambda=0.2$) and $M$ (for example, $M\sim 10^{-3}\text{eV}$).

Using the obtained expansion coefficients \eqref{expan_coeff_k_essence_another}, one can find the dispersion relation  \eqref{disp_rel_gen}.
Due to its cumbersomeness, we do not show it here. % in order not to encumber the paper.

\section{Conclusion}

The process of propagation of a weak GW on the background of DE modeled by different types of matter has been considered.
In doing so, we have used the approximate approach where the scale factor
$a$ has been expanded as a power series at relatively small values of the redshift 
corresponding to the epoch of the present accelerated expansion of the Universe. In this case values of the expansion coefficients
$a_1, a_2$, which are completely determined by parameters of a specific DE model, permit us to find the dispersion relation
\eqref{disp_rel_gen}, whose form depends on the type of DE.

Within the framework of the approximate approach used here, one can find the dispersion relation for various types of DE. As examples
we have considered four kinds of DE (Chaplygin gas, quintessence, two $k$-essence models)
for which the corresponding expansion coefficients have been found.
By changing the values of the free parameters appearing in these expansions, one can find a form of the dispersion relation for DE models used in the literature.
Then, comparing the obtained dispersion relations with observational data that we have at our disposal (or that will be obtained in future experiments),
one can make some conclusions about the viability of various DE models.

\section*{Acknowledgements}
VD and VF gratefully acknowledge support provided by Grant $\Phi.0755$ in fundamental research in natural sciences by the Ministry of Education and Science of Republic of Kazakhstan.
BK and JK gratefully acknowledge support by the German Research Foundation
within the framework of the DFG Research Training Group 1620
{\it Models of gravity}
as well as support by the Volkswagen Stiftung, and support from FP7, 
Marie Curie Actions, People, International Research Staff Exchange
Scheme (IRSES-606096).


\begin{thebibliography}{999}
\bibitem{AmenTsu2010}
 L. Amendola and S. Tsujikawa, {\it  Dark energy: theory and observations} (Cambridge University Press,
Cambridge, England, 2010).

\bibitem{Grishchuk:1974ny}
  L.~P.~Grishchuk,
  %``Amplification of gravitational waves in an istropic universe,''
  Sov.\ Phys.\ JETP {\bf 40}, 409 (1975)
  [Zh.\ Eksp.\ Teor.\ Fiz.\  {\bf 67}, 825 (1974)].
\bibitem{Bernabeu:2011if}
  J.~Bernabeu, D.~Espriu and D.~Puigdomenech,
  %``Gravitational waves in the presence of a cosmological constant,''
  Phys.\ Rev.\ D {\bf 84}, 063523 (2011)
  Erratum: [Phys.\ Rev.\ D {\bf 86}, 069904 (2012)]
  doi:10.1103/PhysRevD.84.063523, 10.1103/PhysRevD.86.069904
  [arXiv:1106.4511 [hep-th]].
\bibitem{Arraut:2012xr}
  I.~Arraut,
  %``About the propagation of the Gravitational Waves in an asymptotically de-Sitter space: Comparing two points of view,''
  Mod.\ Phys.\ Lett.\ A {\bf 28}, 1350019 (2013)
  doi:10.1142/S0217732313500193
  [arXiv:1203.4305 [gr-qc]].
\bibitem{Dunsby:1998hd}
  P.~K.~S.~Dunsby, B.~A.~C.~C.~Bassett and G.~F.~R.~Ellis,
  %``Covariant analysis of gravitational waves in a cosmological context,''
  Class.\ Quant.\ Grav.\  {\bf 14}, 1215 (1997)
  doi:10.1088/0264-9381/14/5/023
  [gr-qc/9811092].
\bibitem{Kleidis:2005}
K. Kleidis and D. B. Papadopoulos
 %EXPLORING THE UNIVERSE ON THE BACK OF A GRAVITATIONAL WAVE.
 COSMOLOGY AND GRAVITATIONAL PHYSICS {\bf 15}, 83 (2005).
 \bibitem{Bento:2003we}
  M.~d.~C.~Bento, O.~Bertolami and A.~A.~Sen,
  %``WMAP constraints on the generalized Chaplygin gas model,''
  Phys.\ Lett.\ B {\bf 575}, 172 (2003)
  doi:10.1016/j.physletb.2003.08.017
  [astro-ph/0303538].
\bibitem{Zlatev:1998tr}
  I.~Zlatev, L.~M.~Wang and P.~J.~Steinhardt,
  %``Quintessence, cosmic coincidence, and the cosmological constant,''
  Phys.\ Rev.\ Lett.\  {\bf 82}, 896 (1999)
  doi:10.1103/PhysRevLett.82.896
  [astro-ph/9807002].
\bibitem{ArkaniHamed:2003uy}
  N.~Arkani-Hamed, H.~C.~Cheng, M.~A.~Luty and S.~Mukohyama,
  %``Ghost condensation and a consistent infrared modification of gravity,''
  JHEP {\bf 0405}, 074 (2004)
  doi:10.1088/1126-6708/2004/05/074
  [hep-th/0312099].
 \end{thebibliography}
\end{document}